\documentclass[11pt,aps,prd,longbibliography,amsmath,amssymb]{revtex4-2}
\usepackage{graphicx}
\usepackage{dcolumn}
\usepackage{bm}
\usepackage{amssymb}
\usepackage{graphics}
\usepackage{slashed}
\usepackage{natbib}
\usepackage{url}
\usepackage[dvipsnames,svgnames,table]{xcolor}
\usepackage[T1]{fontenc}
\usepackage[none]{hyphenat}

\usepackage{diagbox}
\usepackage[normalem]{ulem}

\begin{document}

\title{Qualitative analysis of viscous cosmology}%

\author{Alan G. Cesar}
\email{alangois@cbpf.br}
\author{Mario Novello}
\email{novello@cbpf.br}
\affiliation{Brazilian Center for Research in Physics, R. Dr. Xavier Sigaud, 150, Botafogo, Rio de Janeiro/RJ, Brazil}

\date{\today}

\begin{abstract}

We analyze the effect of higher order viscosity corrections in a spatially homogeneous and isotropic universe. The viscous effects are modeled by introducing a second-order term to the energy density in the equation of state and on the expansion factor. An autonomous dynamical system is then constructed to study the qualitative behavior of the energy density and expansion factor, employing the Raychaudhuri equation and the conservation of energy. We obtain the analytical expressions for the equilibrium points in terms of the model parameters, presenting several phase diagrams. The analysis of these diagrams reveals the presence of bifurcations, which manifest as changes in stability and qualitative behavior. Those changes lead to different configurations for the universe, such as bouncing, rebouncing and cyclic scenarios.

\end{abstract}

\maketitle

\section{Introduction}

The standard model for cosmology, $\Lambda$CDM, shows several successes in explaining cosmological observational data. Some examples include the cosmic microwave background (CMB), large-scale structure, and the apparent accelerated expansion of the universe. Nonetheless, it faces challenges when different observations are taken into consideration \cite{Ellis2018}. Some of these problems show up when considering the nature of dark matter and dark energy \cite{Giare2025,des2025}, the Hubble tension \cite{Valentino2021,Hu2023}, the presence of the Big-Bang singularity, and the horizon and flatness problems.

Taking these issues into consideration, there is a significant presence in the literature of discussions on alternative cosmological models that attempt to address some or all of these questions. The singularity issues are discussed, for example, within bouncing \cite{Novello2008,Brandenberger2017,Novello1979} and cyclic models \cite{Steinhardt2002,Baum2007,Novello2009,Medeiros2012}, while maintaining homogeneity and isotropy. Other models seek to change the geometric structure of the universe altogether—the so-called modified theories of gravity, such as $f(R)$ gravity \cite{Sotiriou2010,Nojiri2006,Nojiri2017,Nojiri2011}. In particular, we are interested in models that do not modify the general relativity (GR) framework, but rather the matter that sources it. This approach is commonly realized with scalar fields \cite{Nojiri2006,BenDayan2019,Giacomini2020}, quantum fields \cite{Sakellariadou2017,Pintoneto2012,Bishop2023}, or non-linear electrodynamics \cite{Lorenci2002,Novello2004,Novello2012,Breton2010,Ovgun2018}.

In this paper, we analyze the effect of nonlinear viscous processes by introducing second-order corrections into the equation of state (EoS) of the fluid that sources the cosmological model, a procedure that was considered by Zel'dovich as a consequence of matter creation~\cite{Zel}. It has been used as a source for cosmological models before~\cite{novello1,novello1b,novello2,Novello1993b,Aguilar-Perez2022}. The results in this work are an extension of the qualitative results obtained by one of the authors in~\cite{Novello1980,novello2}, with a focus on the new behavior that emerges from a second-order correction in $\theta$. We provide further details about the constraints on the equilibria and their stability, an analysis of the non-hyperbolic origin, and  show the presence of bifurcations, as discussed in the theory of non-linear autonomous dynamical systems~\cite{kuznetsov2004}.

The paper is organized as follows: First, we present the background of the dynamical system construction in Section~II, followed by its application in Section~III. Section~III is divided into four subsections: the general case is discussed in Subsection~A, and particular cases are addressed in Subsections~B and~C. Lastly, we discuss the stability of the non-hyperbolic equilibrium in Subsection~D.

\section{The dynamical system of cosmology}

Let us consider a homogeneous and isotropic scenario described by the metric  
\begin{equation}
\label{metric}
   ds^2 = dt^2 - a^2(t)\left[d\chi^2+f^2(\chi)\left(d\theta^2 +\sin^2(\theta)d\phi^2\right)\right],
\end{equation}
which satisfies the equations of GR for a fluid with energy density $\rho$ and pressure $p$. The equations of GR reduce to the following pair:
\begin{equation}
\label{continuity}
  	\dot{\rho} + \left( \rho + p \right) \, \theta = 0,
\end{equation}
\begin{equation}
\label{rachayduri}
  	\dot{\theta} + \frac{\theta^2}{3} + \frac{1}{2} \, \left( \rho + 3p \right) = 0,
\end{equation}
where $\theta = 3 \dot{a}/a$ is the expansion factor. We analyze the corresponding planar dynamical system, described by the set
\begin{equation*}
    \dot{\rho} = F(\rho, \theta), \qquad \dot{\theta} = H(\rho, \theta).
\end{equation*}
In the following sections, we present a qualitative analysis of the phase diagrams obtained from the introduction of different viscous processes into the EoS.

\section{A cosmology with quadratic viscosity corrections}

The EoS includes quadratic corrections to the energy density and expansion factor, and can be written as follows:
\begin{equation}
    \label{generalEoS}
    p = \lambda \rho + \sigma\rho^2+\alpha\theta+\beta\theta^2,
\end{equation}
where $\lambda$, $\sigma$, $\alpha$, and $\beta$ are free parameters. The parameter $\lambda$ will be discussed within the interval $-1<\lambda<1$, partly due to the expected behavior of standard cosmology under specific choices of this parameter in the linear EoS (\cite{Ellis2018}), and partly due to the upper limit on the speed of sound (\cite{novello1}). The $\sigma$ and $\alpha$ parameters are small, and the presence of this type of correction has already been discussed in the literature (\cite{novello1b,novello2,Novello1980}). The quadratic correction in $\theta$ has previously been considered in inhomogeneous cosmology (\cite{novello1}), and briefly in a homogeneous background (\cite{Novello1980}), constraints on its values can also be discussed through a careful analysis of the observational data (\cite{Koussour2024,Odintsov2020,Villalobos2025}).  We will begin by discussing the general case, and then proceed to the particular cases where $\alpha=0$ or $\sigma=0$. At the end of this section, we will briefly discuss the stability of the origin equilibrium.

\subsection{The general case}

From equations (\ref{continuity}) and (\ref{rachayduri}), the following dynamical system is obtained:
\begin{equation}
\label{generalds}
    \dot{\rho} = -\theta\left((1+\lambda)\rho+\sigma\rho^2+\alpha\theta+\beta\theta^2\right),\qquad \dot{\theta}=-\frac{\theta^2}{3}-\frac{1}{2}\left((1+3\lambda)\rho+3\sigma\rho^2+3\alpha\theta+3\beta\theta^2\right).
\end{equation}
The fixed points can be obtained when $\theta=0$ and when $(1+\lambda)\rho+\sigma\rho^2+\alpha\theta+\beta\theta^2=0$. When the expansion factor is zero, the fixed points are:
\begin{equation}
    \label{nullexpansionequilibria}
    P_0 = (0,0) \quad \mathrm{and} \qquad P_1=  \left(\frac{-1-3\lambda}{3\sigma},0\right).
\end{equation}
Note that $P_1$ is only valid when $\sigma\neq 0$; the situation in which $\sigma=0$ must be discussed separately, as viscosity becomes more relevant than the existence of a maximum density. The other candidate fixed points require a more careful analysis. Solving for $\rho=0$ when $\theta\neq 0$ leads to the following expression:
\begin{equation}
    \label{rhopm}
    \rho_{\pm} = \frac{-1-\lambda}{2\sigma}\pm\frac{1}{2\sigma}\sqrt{(1+\lambda)^2-4\sigma(\alpha\theta+\beta\theta^2)}.
\end{equation}
By substituting into the $\dot{\theta}$ expression and performing algebraic manipulations, the following result is obtained for $\theta\neq 0$:
\begin{equation}
    \label{cardanogeral}
    \theta^3+\frac{3\theta}{\sigma}\left(1+\lambda+3\beta\right)+\frac{9\alpha}{\sigma}=0.
\end{equation}
From equation (\ref{cardanogeral}) it is possible to obtain up to three candidate fixed points, which must then be substituted into equation (\ref{rhopm}) and subsequently verified using equation (\ref{generalds}). The discriminant of this depressed cubic then provides a way to discuss the existence of the candidate solutions. Note that
\begin{equation}
    \label{discriminantgeral}
    \Delta=\frac{81 \alpha^2}{4\sigma^2}+\left(\frac{1+\lambda+3\beta}{\sigma}\right)^3,
\end{equation}
where, $\Delta$ is the discriminant of the depressed cubic. When $\Delta\leq 0$, equation (\ref{cardanogeral}) has three real solutions, with $\Delta=0$ corresponding to two repeated roots. When $\Delta>0$, there is only one real root. Each real root can be substituted into equation (\ref{rhopm}) to obtain a pair of candidates. Note that only one of these corresponds to an actual fixed point of the dynamical system.

From equation (\ref{discriminantgeral}), it is apparent that the sign of $\alpha$ is not relevant to the existence of the roots. However, its absolute value does affect the range of the parameters that define the regions where $\Delta$ has different signs. Additionally, these regions are separated by the line given by $\frac{1+\lambda+3\beta}{\sigma}$. When this ratio is positive, $\Delta$ is always positive. Nevertheless, for any given value of $\sigma$, it is always possible to choose values of $\beta$ and $\lambda$ such that the sign of $\Delta$ changes.

Before delving deeper into the existence of these equilibria, we will first discuss the stability of $P_0$ and $P_1$, since their algebraic expressions are more manageable. A preliminary discussion on the equilibria stability can be carried out by linearizing the dynamical system in the neighborhood of each equilibrium point. Unfortunately, the stability of the origin equilibrium $(P_0)$ cannot be determined using this approach, since the Jacobian evaluated at $P_0$ has eigenvalues $r=0$ and $r=\frac{-3\alpha}{2}$. Therefore, it is necessary to examine the effects of the non-linear terms to study the stability of this equilibrium, this will be discussed in subsection D.

For $P_1$, the eigenvalues are given by the pair:
\begin{equation}
    \label{eingenP1}
    r_{\pm}=-\frac{3\alpha}{4}\pm\frac{1}{2}\sqrt{\frac{9\alpha^2}{4}+\frac{4(1+3\lambda)^2}{9\sigma}},
\end{equation}
Since neither eigenvalue is zero, the stability of the equilibrium can be fully determined using equation (\ref{eingenP1}). From equation (\ref{eingenP1}), the sign of the discriminant determines whether the equilibrium behaves as a focus/source, as a saddle point, or as an unstable/stable node. The only transition happens when $\sigma$ changes sign. Note that for $\sigma>0$ the equilibrium always behaves as a saddle, while for $\sigma<0$ there is the transition from a focus or source to an unstable/stable node within a small region that depends on the fixed values of all three parameters. Furthermore, for $\lambda=-\frac{1}{3}$—a value commonly regarded as the upper limit for a "dark energy cosmology" (\cite{Ellis2018})—the eigenvalues become zero, which is consistent with the degeneracy of this equilibrium with the origin. The sign of $\alpha$ here determines the stability of the equilibrium when it behaves as a source ($\alpha<0$) or as a focus ($\alpha>0$). For $\alpha=0$, centers exist; this case will be discussed in more detail in its own section. It is also important to point out that neither the existence nor the stability of this equilibrium depends on the parameter $\beta$. The relevance of this parameter appears in the stability analysis of the equilibria candidates given by equations (\ref{cardanogeral}) and (\ref{rhopm}).

\begin{figure}[h]
  \centering
  \includegraphics[width=1\textwidth]{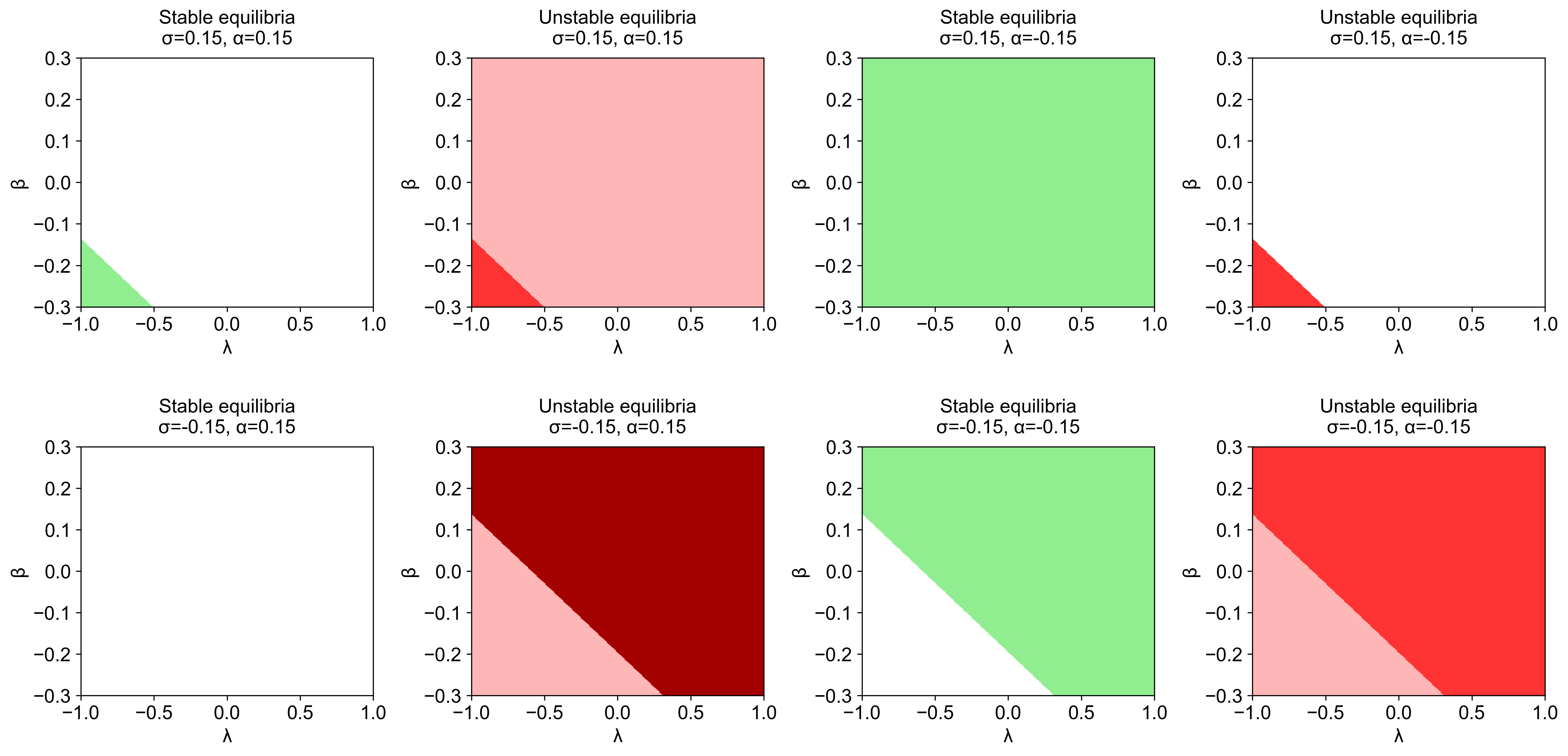} 
  \caption{The possible combinations of $\sigma=\pm0.15$ and $\alpha=\pm0.15$ are presented. Light green regions indicate the existence of a single stable equilibrium, while light pink regions represent a single unstable equilibrium. The red region corresponds to the existence of two unstable equilibria, and the dark red region to three unstable equilibria. White regions indicate that no stable or unstable equilibria exist.}
  \label{regioncardanogeral}
\end{figure}

The other candidates for equilibria have a complex algebraic form. To evaluate the existence of the fixed point and their stability, we construct region graphs over a range of $\lambda$ and $\beta$ values, for fixed choices of $\alpha$ and $\sigma$. The regions can be observed in Fig.~\ref{regioncardanogeral}.

Fig.~\ref{regioncardanogeral} shows, for a given choice of $\alpha$ and $\sigma$, the number and stability of equilibria, as determined by linearizing the dynamical system. It can be seen that, for most parameter combinations, stable and unstable equilibria coexist for particular values of $\lambda$ and $\beta$, as exemplified by the small regions in the graphs for $\sigma=0.15$ and $\alpha=0.15$. The change in stability of equilibria as either $\alpha$ or $\sigma$ changes sign is also evident. When $\sigma=0.15$, there is a transition from one stable equilibrium to two unstable equilibria as $\alpha$ changes sign, while a stable equilibrium now exists throughout the entire range of parameters considered. For negative $\sigma$ and positive $\alpha$, stable equilibria disappear, but when $\alpha$ also becomes negative, one of the three unstable equilibria becomes stable. The graphs were generated for specific absolute values of $\sigma$ and $\alpha$, but changing these values does not affect the qualitative behavior of the regions. Therefore, we focus on the effects of changing their signs. 

It is also evident from Fig.~\ref{regioncardanogeral} that varying the values of $\beta$ and $\lambda$ alters the qualitative behavior of the fixed points. The lines separating the regions indicate the appearance of new equilibria or a change in their stability. From the graphs on the top row, note that for $\lambda=-1$ and very negative $\beta$, there are two unstable equilibria and one stable equilibrium when $\alpha$ is positive. However, higher values of $\lambda$ or $\beta$ result in a single unstable equilibrium and no stable equilibrium. Similarly, in the bottom row, it is possible to observe the emergence of two additional unstable equilibria when larger values of $\lambda$ or $\beta$ are considered.

The three equilibria given by equation (\ref{cardanogeral}) exhibit various behaviors, including attractive and repulsive nodes as well as saddle points. When $\Delta>0$ in equation (\ref{discriminantgeral}) and there is only one equilibrium, this fixed point behaves as an attractive node ($\alpha<0$) or a repulsive node ($\alpha>0$) when $\sigma>0$, or as a saddle when $\sigma<0$. When $\Delta<0$ and three equilibria appear, two of them always behave as saddles, while the third acts as a repulsive node ($\alpha>0$) or an attractive node ($\alpha<0$). It becomes evident that $\alpha$ and $\sigma$ have direct impact on the equilibria stability, while $\lambda$ and $\beta$ control the number of equilibria and their qualitative behavior. For a less mathematical discussion and to explore the cosmological consequences of the corrections, it is necessary to study the phase diagrams of the dynamical system.

\begin{figure}[h]
  \centering
  \includegraphics[width=1\textwidth]{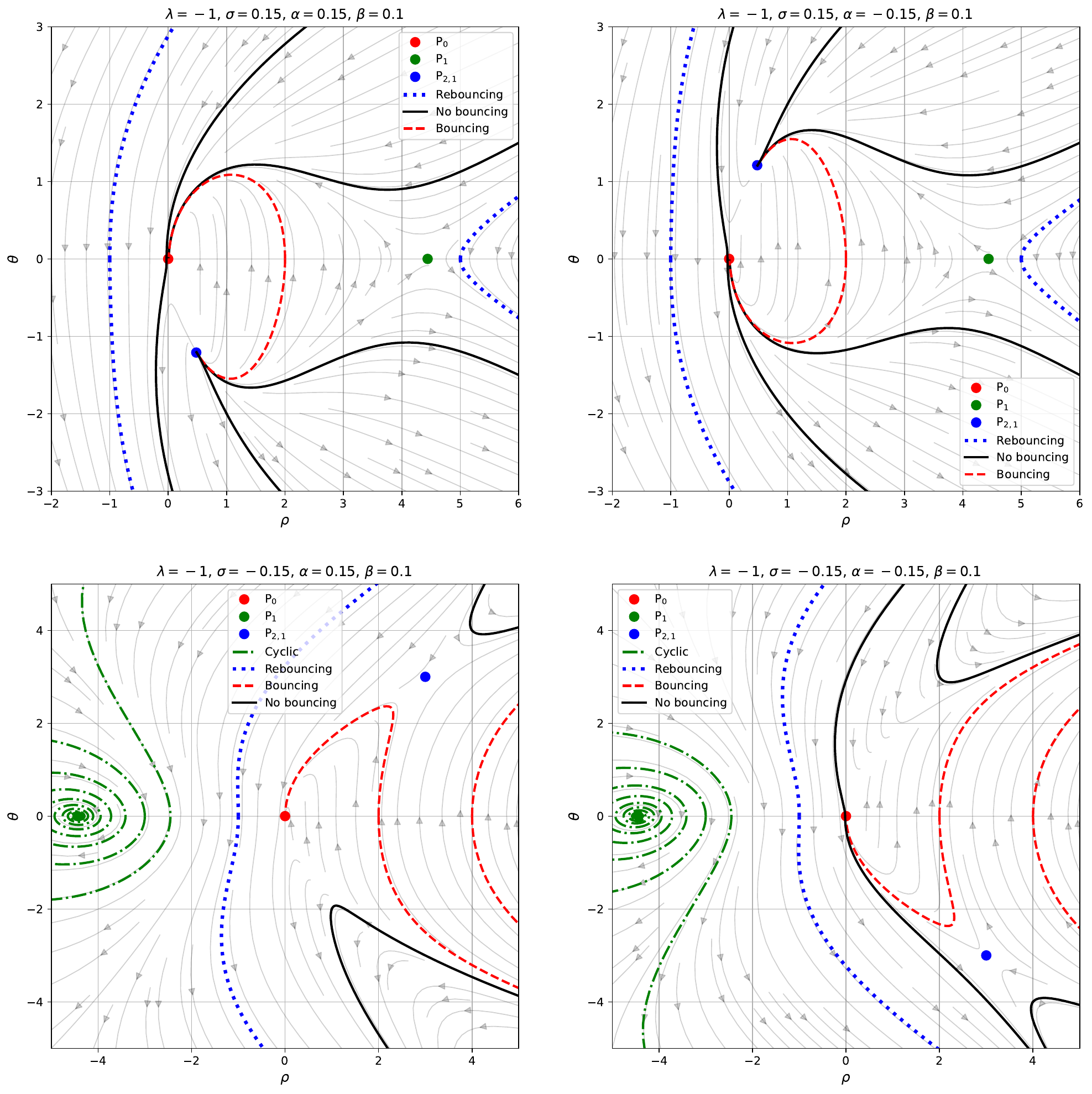} 
  \caption{Phase diagrams with $\lambda=-1$ and $\beta=0.1$, showing the different combinations of $\sigma=\pm0.15$ and $\alpha=\pm0.15$. The red dashed curves are bouncing solutions, the blue dotted curves indicate rebouncing solutions, the green dot-dashed curves correspond to cyclic solutions, and the black solid curves are those solutions that do not exhibit a bounce.}
  \label{gerallambda1n}
\end{figure}

Fig.~\ref{gerallambda1n} presents the choice $\lambda=-1$ and $\beta=0.1$ considering the combinations $\sigma=\pm0.15$ and $\alpha=\pm0.15$.

Fig.~\ref{gerallambda1n} clearly illustrates how the qualitative behavior of $P_1$ depends on the sign of $\sigma$, and in parallel, how stability is determined by the sign of $\alpha$. Among the three equilibria from equation~(\ref{cardanogeral}), one is always present, with its characteristics evolving accordingly. For $\sigma>0$, $P_1$ acts as a saddle point, while $P_{2,1}$ is a node. Here, $P_1$ forms the upper boundary of a region where solutions begin in a collapsing phase before transitioning to accelerated expansion—a phenomenon known as a bounce, which occurs at the point of maximum positive energy density. These bouncing trajectories can originate from either the origin, $P_0$, when $\alpha<0$, or from the repulsive node,  $P_{2,1}$, when $\alpha>0$, reflecting the stability properties of these fixed points.

In particular, for $\sigma>0$ and $\alpha>0$, within a "cosmological constant" universe (quotation marks indicate the approximate nature due to corrections in the EoS) with $\lambda=-1$, a universe starting in a decelerated collapse may bounce into a static configuration after passing through accelerated and then decelerated expansion. However, if the collapse is too rapid or the initial energy too low, collapse continues indefinitely—a result of the instability at $P_{2,1}$ and the semi-stable character of $P_0$. For high energy densities, solutions may initially expand, reach a maximum size and minimum energy density, then collapse forever. If the trajectory enters the negative energy density regime, it also leads to endless collapse — we are calling those rebouncing solutions. Other trajectories may approach the static universe if starting in expansion, or else collapse indefinitely.

With $\alpha<0$, the instability at the origin increases, and bouncing solutions asymptotically approach a fixed point with positive energy density and constant expansion. When $\sigma$ becomes negative, $P_1$ turns into a focus for $\alpha>0$, or a source for $\alpha<0$, leading to oscillatory solutions around a fixed negative energy density. Bouncing solutions remain, now bounded by $P_{2,1}$, which acts as a saddle. The sign of $\alpha$ continues to determine the fate of these solutions: for $\alpha>0$, they may approach a static universe with zero energy density, while for $\alpha<0$, they undergo indefinite accelerated expansion. There is also a small region with rebouncing solutions, and in this regime, bouncing solutions can also occur at higher energy densities.

\begin{figure}[h]
  \centering
  \includegraphics[width=1\textwidth]{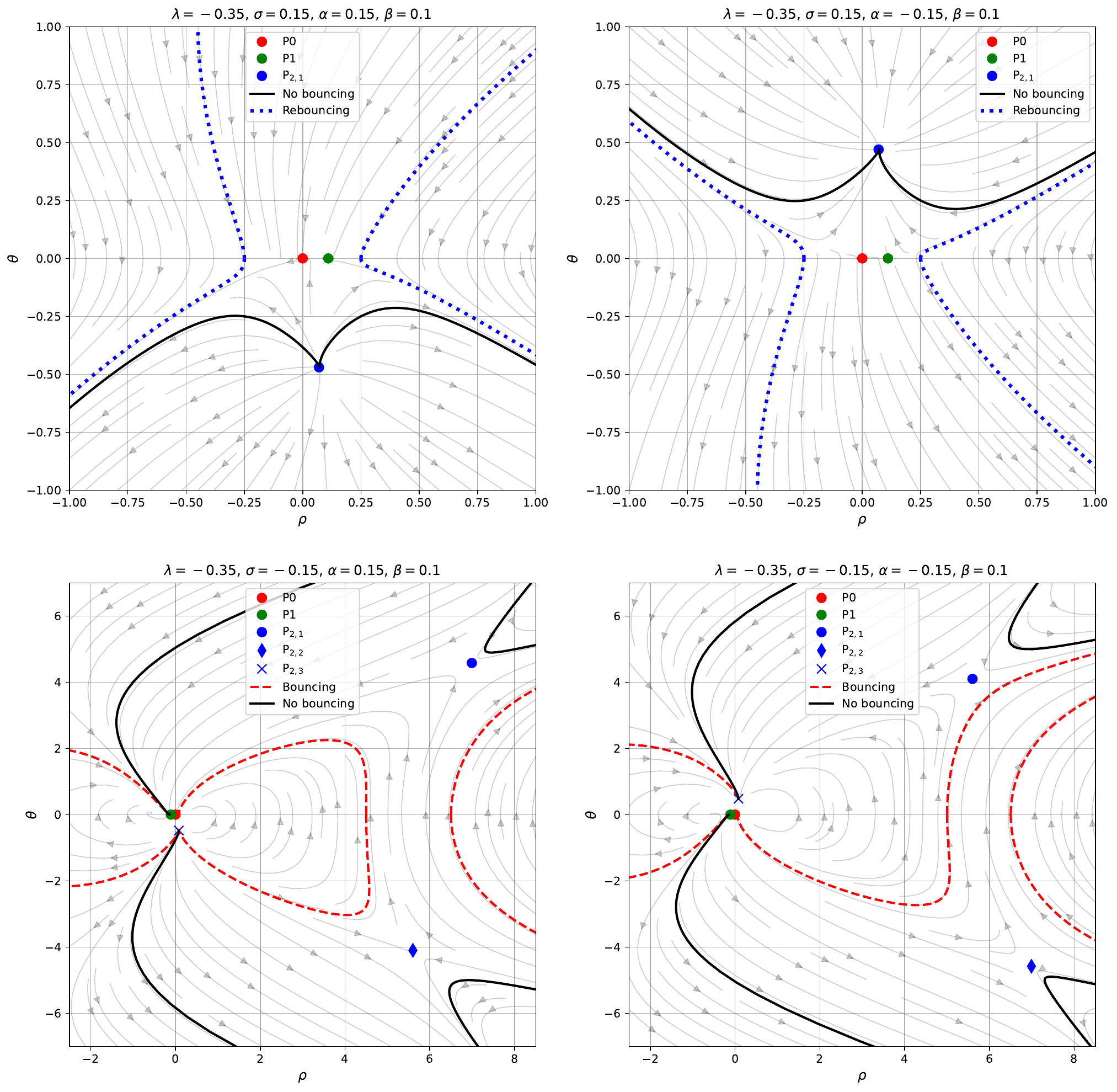} 
  \caption{Phase diagrams with $\lambda=-0.35$ and $\beta=0.1$, showing the different combinations of $\sigma=\pm0.15$ and $\alpha=\pm0.15$. The red dashed curves are bouncing solutions, the blue dotted curves indicate rebouncing solutions, and the black solid curves are those solutions that do not exhibit a bounce.}
  \label{gerallambda0.35nb0.1p}
\end{figure}

\begin{figure}[h]
  \centering
  \includegraphics[width=1\textwidth]{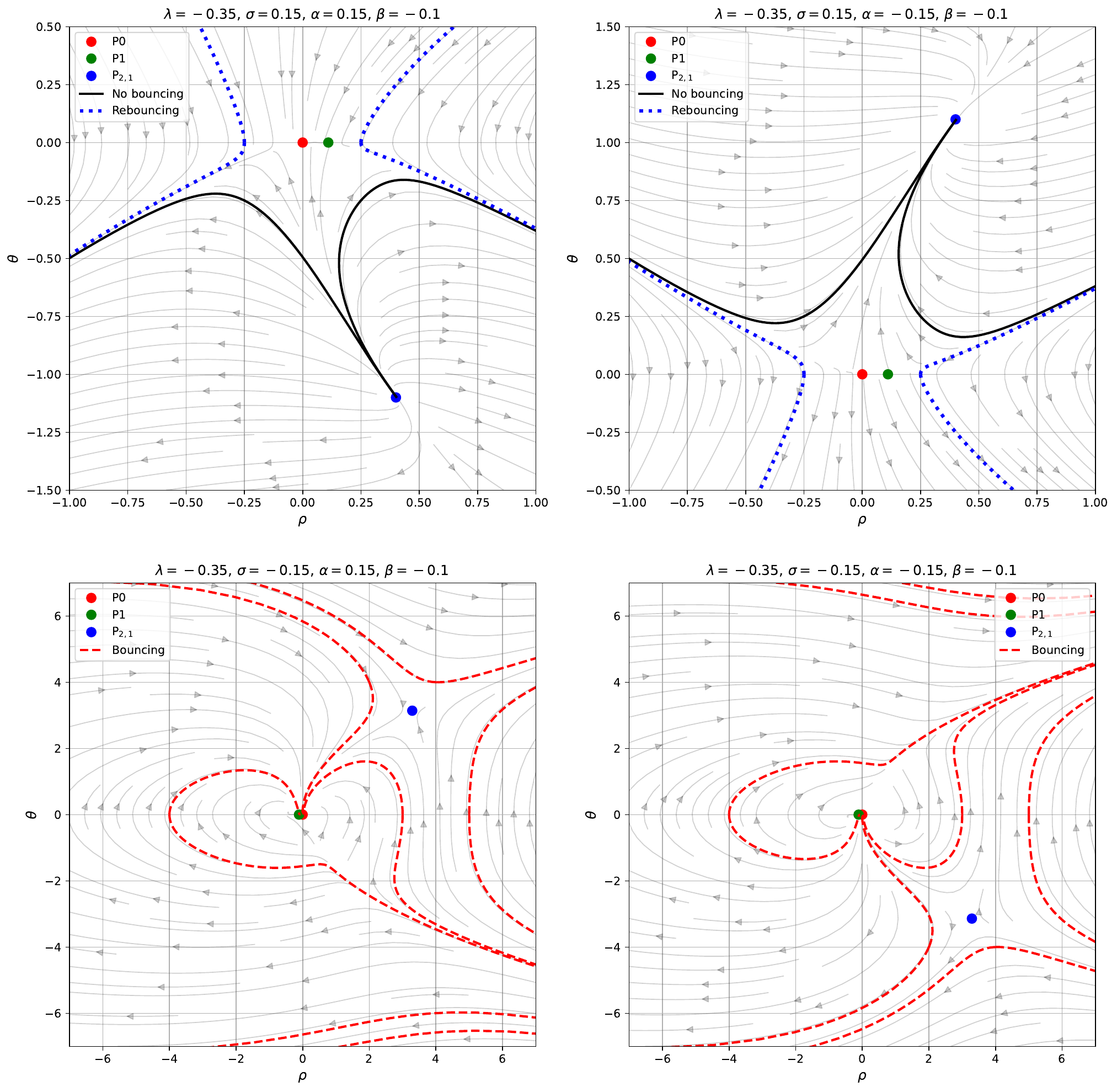} 
  \caption{Phase diagrams with $\lambda=-0.35$ and $\beta=-0.1$, showing the different combinations of $\sigma=\pm0.15$ and $\alpha=\pm0.15$. The red dashed curves are bouncing solutions, the blue dotted curves indicate rebouncing solutions, and the black solid curves are those solutions that do not exhibit a bounce.}
  \label{gerallambda0.35nb0.1n}
\end{figure}

Fig.~\ref{gerallambda0.35nb0.1p} and ~\ref{gerallambda0.35nb0.1n} illustrate the dynamics for $\lambda=-0.35$ and $\beta=\pm0.1$. In Figure~\ref{gerallambda0.35nb0.1p}, observe the behavior of $P_{2,1}$, which acts as an unstable node for $\alpha>0$ or a stable node for $\alpha<0$ when $\sigma>0$. Meanwhile, $P_1$ consistently behaves as a saddle point, as expected from equation ~(\ref{eingenP1}). Collapse solutions originating from $P_{2,1}$ will undergo perpetual collapse; however, if the collapse is sufficiently decelerated, these trajectories may eventually be redirected by the tendency flow generated by $P_0$ and $P_1$ for $\alpha>0$. A similar scenario occurs for $\alpha<0$, but in this case $P_{2,1}$ represents a final state for a universe undergoing constant-rate accelerated expansion. But, due to its asymmetry it is possible for a bouncing solution to expand forever ($\alpha<0$) or result from a quick collapse ($\alpha>0$).

For $\sigma<0$, the qualitative behavior changes significantly. The equilibria $P_{2,2}$ and $P_{2,3}$ emerge, functioning as a saddle point and as either an unstable node ($\alpha>0$) or stable node ($\alpha<0$), respectively. Additionally, $P_1$ now mimics $P_{2,3}$, but with reversed stability. In both signs of $\alpha$, a region of bouncing solutions appears at higher energy densities, alongside a secondary bouncing region defined by $P_0$, $P_1$, and $P_{2,3}$. These trajectories either originate from $P_{2,3}$ when $\alpha>0$, or approach $P_{2,3}$ when $\alpha<0$. Increasing $|\theta|$ or decreasing $|\rho|$ leads to solutions that collapse indefinitely if they start at $P_{2,3}$, or that collapse until reaching $P_0$ or $P_1$ for $\alpha>0$. The inverse applies for $\alpha<0$, due to the swapped stability of the equilibria. The saddle points $P_{2,1}$ and $P_{2,2}$ inhibit bouncing for solutions with both high energy density and high expansion parameter, resulting in eternal collapse or expansion. When $\beta<0$ (see Fig.~\ref{gerallambda0.35nb0.1n}), the behavior for $\sigma>0$ is unchanged. However, for $\sigma<0$, only $P_{2,1}$ remains; the source or sink associated with $P_{2,3}$ now originates from infinity, and the absence of $P_{2,2}$ enables all collapsing solutions to eventually bounce into an expanding universe.

\begin{figure}[h]
  \centering
  \includegraphics[width=1\textwidth]{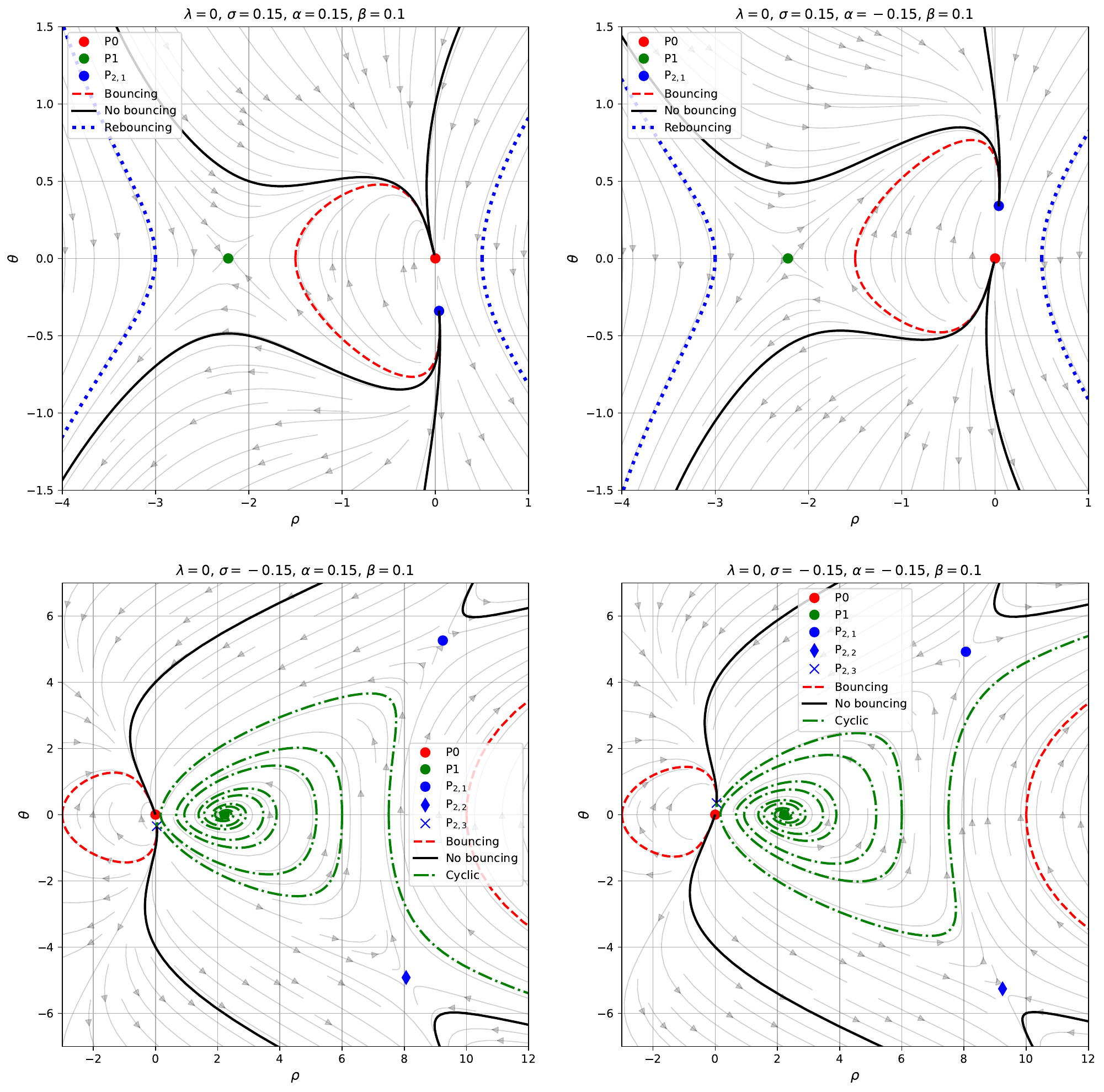} 
  \caption{Phase diagrams with $\lambda=0$ and $\beta=0.1$, showing the different combinations of $\sigma=\pm0.15$ and $\alpha=\pm0.15$. The red dashed curves are bouncing solutions, the blue dotted curves indicate rebouncing solutions, the green dot-dashed curves correspond to cyclic solutions, and the black solid curves are those solutions that do not exhibit a bounce.}
  \label{gerallambda0b0.1p}
\end{figure}

\begin{figure}[h]
  \centering
  \includegraphics[width=1\textwidth]{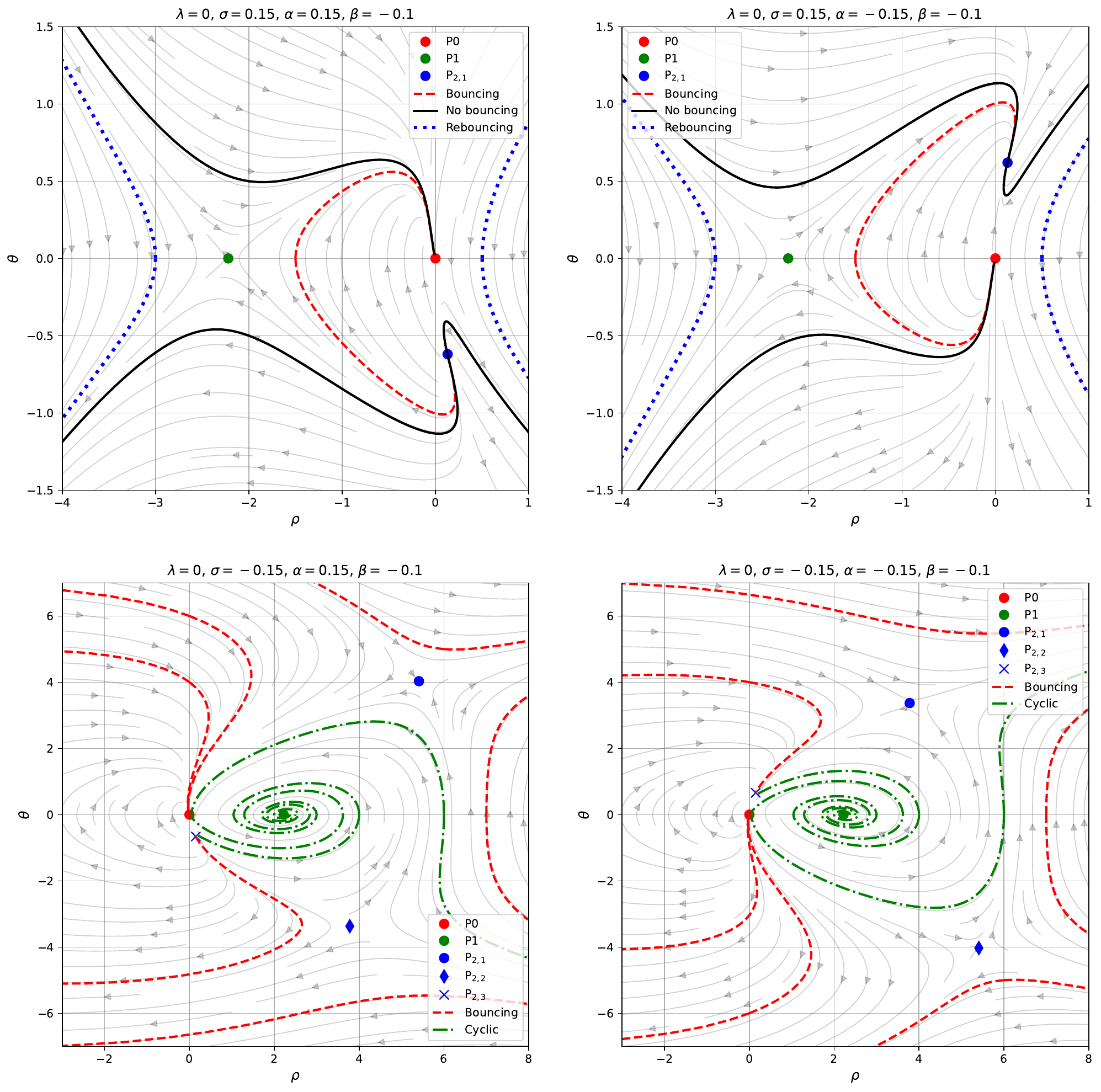} 
  \caption{Phase diagrams with $\lambda=0$ and $\beta=-0.1$, showing the different combinations of $\sigma=\pm0.15$ and $\alpha=\pm0.15$. The red dashed curves are bouncing solutions, the blue dotted curves indicate rebouncing solutions, the green dot-dashed curves correspond to cyclic solutions, and the black solid curves are those solutions that do not exhibit a bounce.}
  \label{gerallambda0b0.1n}
\end{figure}

Fig.~\ref{gerallambda0b0.1p} and~\ref{gerallambda0b0.1n} show the case where $\lambda=0$ and $\beta=\pm0.1$. This value of $\lambda$ is commonly associated with a universe in which a dust-like fluid is the main component.

When $\beta>0$, as depicted in Fig.~\ref{gerallambda0b0.1p}, the overall behavior closely resembles that of Fig.~\ref{gerallambda1n} for $\sigma>0$, with the main difference being that the saddle point $P_1$ now occurs at negative energy density. Consequently, the region of bouncing solutions appears for negative densities. For $\sigma<0$, $P_1$ becomes a focus when $\alpha>0$ and a source when $\alpha<0$, giving rise to two types of solutions oscillating around this equilibrium. In the case $\alpha>0$, solutions either originate from $P_{2,3}$ and settle into a static universe with positive energy density, $P_1$, or emerge from a decelerated collapse. For $\alpha<0$, solutions start from a static universe with positive energy density and evolve either toward a universe with constant expansion and very low energy density, $P_{2,3}$, or toward accelerated expansion with increasing energy density. Solutions that emerge from $P_{2,3}$ can also bounce back into a static empty universe, $P_0$, when the energy density becomes negative, or collapse forever. Solutions will also reach $P_0$ if they start from decelerated expansion initial conditions.

The saddle points $P_{2,1}$ and $P_{2,2}$ delineate the bouncing region, separating it from eternally collapsing or expanding universes. However, their asymmetry permits cyclic solutions to transition between collapsing and expanding phases. Changing the sign of $\beta$
(see Figure~\ref{gerallambda0b0.1n}) has an analogue behavior as seen before allowing the changing in the orientation of the solution curves for $\sigma<0$, enabling all collapsing trajectories to eventually bounce into expansion. Other noteworthy cases include $\lambda=-\frac{1}{3}$, where $P_1$ and $P_0$ merge; the resulting dynamics are analogous to those found for $\lambda=-0.35$. Additionally, for all $\lambda>0$, the behavior matches that described for $\lambda=0$. 

We can summarize the effect of the parameter $\beta$ as follows: it provides a second parametric constraint when studying the presence of bifurcations and the stability of equilibrium points, and it changes the flow of the solution curves when it changes sign, allowing regions of collapsing solutions to bounce into accelerated expansion phases. The latter effect appears when $\sigma < 0$, and oscillatory behavior around $P_1$ can be observed. Furthermore, the freedom to choose the value of $\beta$ allows for different qualitative behaviors of the solutions for various values of $\lambda$, as shown in Figure~\ref{regioncardanogeral}. 

In the following sections, we address special cases where the analytic results above do not apply, specifically for $\sigma = 0$ and $\alpha = 0$. The case $\beta = 0$ has already been discussed in the literature~\cite{Saadat2013, Colistete2007, Eckart1940, Weinberg1972, Mathew2014}.

\subsection{A case where energy density is not maximized}

This section studies the case in which $\sigma=0$. The first noticeable effect is the disappearance of one of the fixed points; now, the dynamics evolve around the following equilibria:

\begin{equation} 
\label{fixedps0} 
P_{0} = (0,0), \qquad P_1 = \left(\frac{3\alpha^2}{(1+\lambda+3\beta)^2},-\frac{3\alpha}{1+\lambda+3\beta}\right). 
\end{equation}

\begin{figure}[h] 
\centering 
\includegraphics[width=1\textwidth]{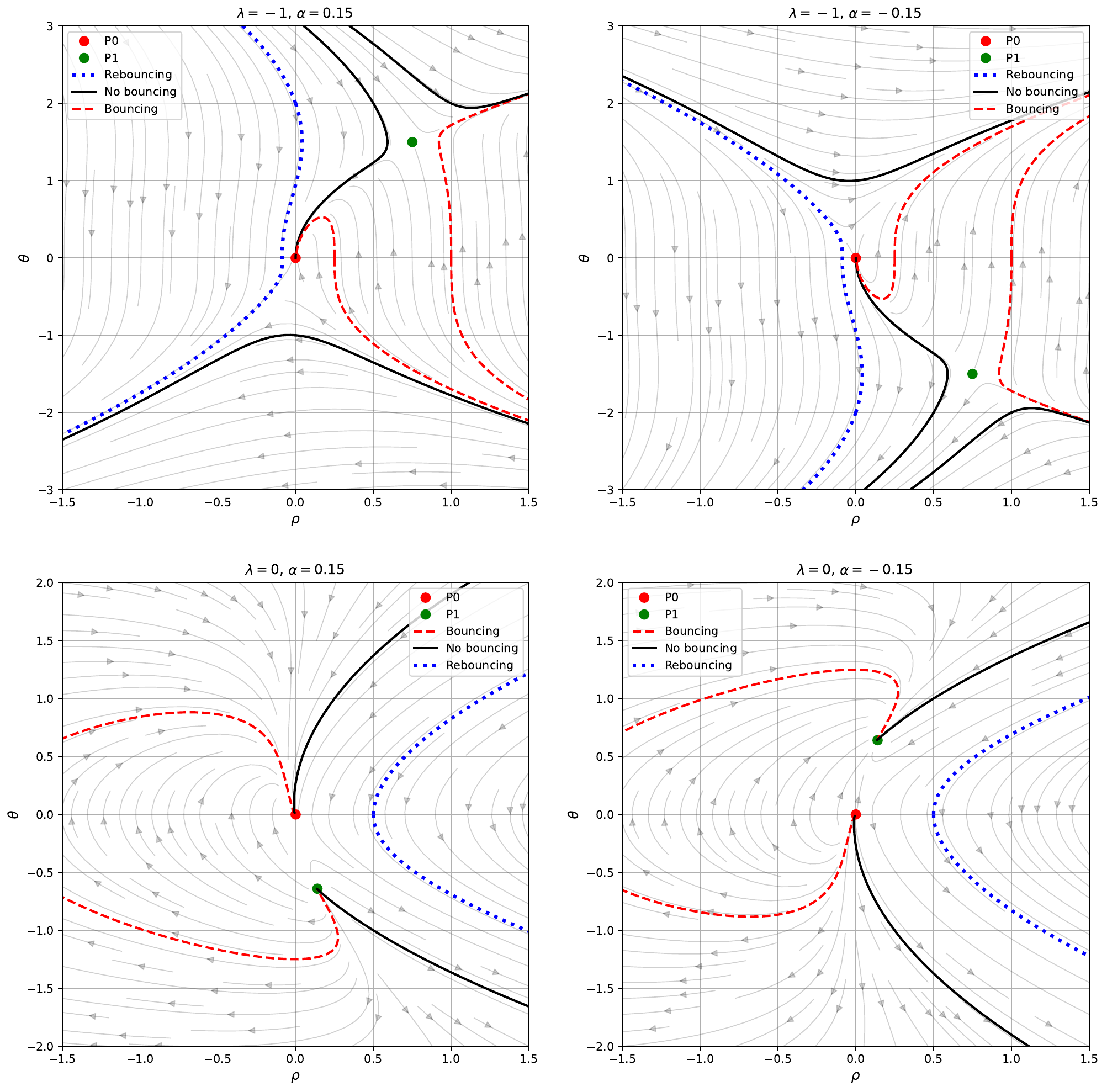} 
\caption{Phase diagrams with $\sigma=0$, $\lambda=-1$ (top row) and $\lambda=0$ (bottom row). All figures are for $\beta=-0.1$, showing the choices of $\alpha=\pm0.15$. The red dashed curves represent bouncing solutions, the blue dotted curves indicate rebouncing solutions, the green dot-dashed curves correspond to cyclic solutions, and the black solid curves correspond to solutions that do not exhibit a bounce.} 
\label{sigma0case} 
\end{figure}

The fixed point $P_1$ in equation~(\ref{fixedps0}) is only valid for $1+\lambda+3\beta\neq 0$. For these choices of parameters, and for $\alpha=0$, the equilibria degenerate with the origin, resulting in only the origin as a fixed point. When $1+\lambda+3\beta>0$, we have either a stable node ($\alpha<0$) or an unstable node ($\alpha>0$). For $1+\lambda+3\beta<0$, $P_1$ always behaves as a saddle. Figure~\ref{sigma0case} shows the different possible scenarios for this dynamical system.

The fixed point $P_1$ behaves similarly to one of the equilibria described by $P_{2,i}$ in the previous section, delimiting bouncing regions and regions that only collapse or expand. The value of $\alpha$ flips the phase diagram vertically, changing the stability of the nodes. Notice that only when $P_1$ represents an expanding universe does it behave as a stable node; its collapsing form is always unstable. When $P_1$ behaves as a node, the regions where bouncing solutions occur change to regions for rebouncing solutions. When $\lambda=-1$ and $\alpha>0$ the saddle given by $P_1$ separate the bounce solution that reach $P_0$ from those that forever expand with an accelerated rate. When $\alpha<0$, $P_0$ functions as source from these bouncing solutions, having the same fate as the collapsing solutions with higher energy densities.

\subsection{The absence of linear viscosity}

\begin{figure}[h] 
\centering 
\includegraphics[width=1\textwidth]{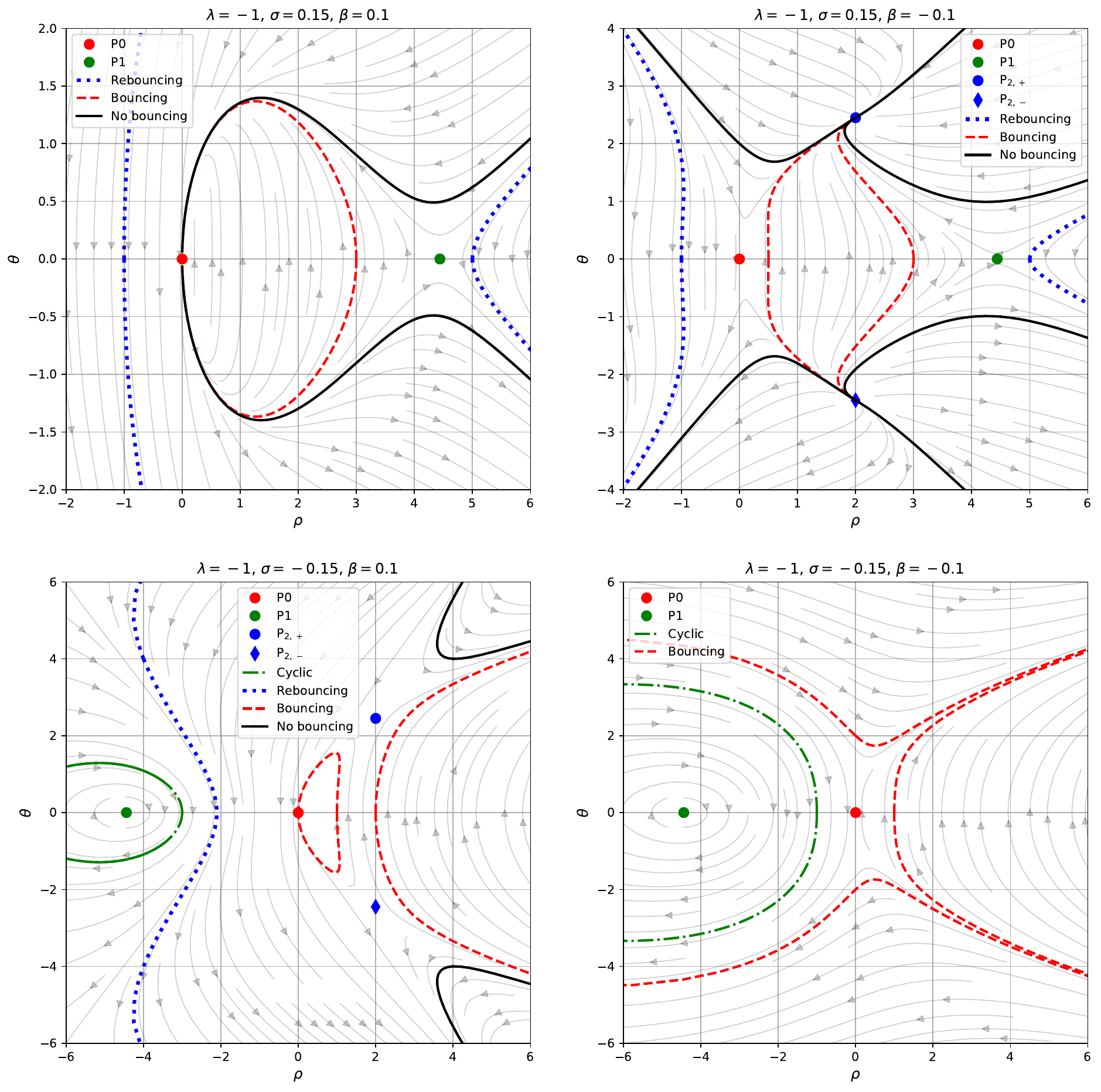} 
\caption{Phase diagrams with $\alpha=0$ and $\lambda=-1$. All figures correspond to the combinations $\beta=\pm0.1$ and $\sigma=\pm0.15$. The red dashed curves represent bouncing solutions, the blue dotted curves indicate rebouncing solutions, the green dot-dashed curves correspond to cyclic solutions, and the black solid curves correspond to solutions that do not exhibit a bounce.} 
\label{alpga0lambda1n} 
\end{figure}

\begin{figure}[h] 
\centering 
\includegraphics[width=1\textwidth]{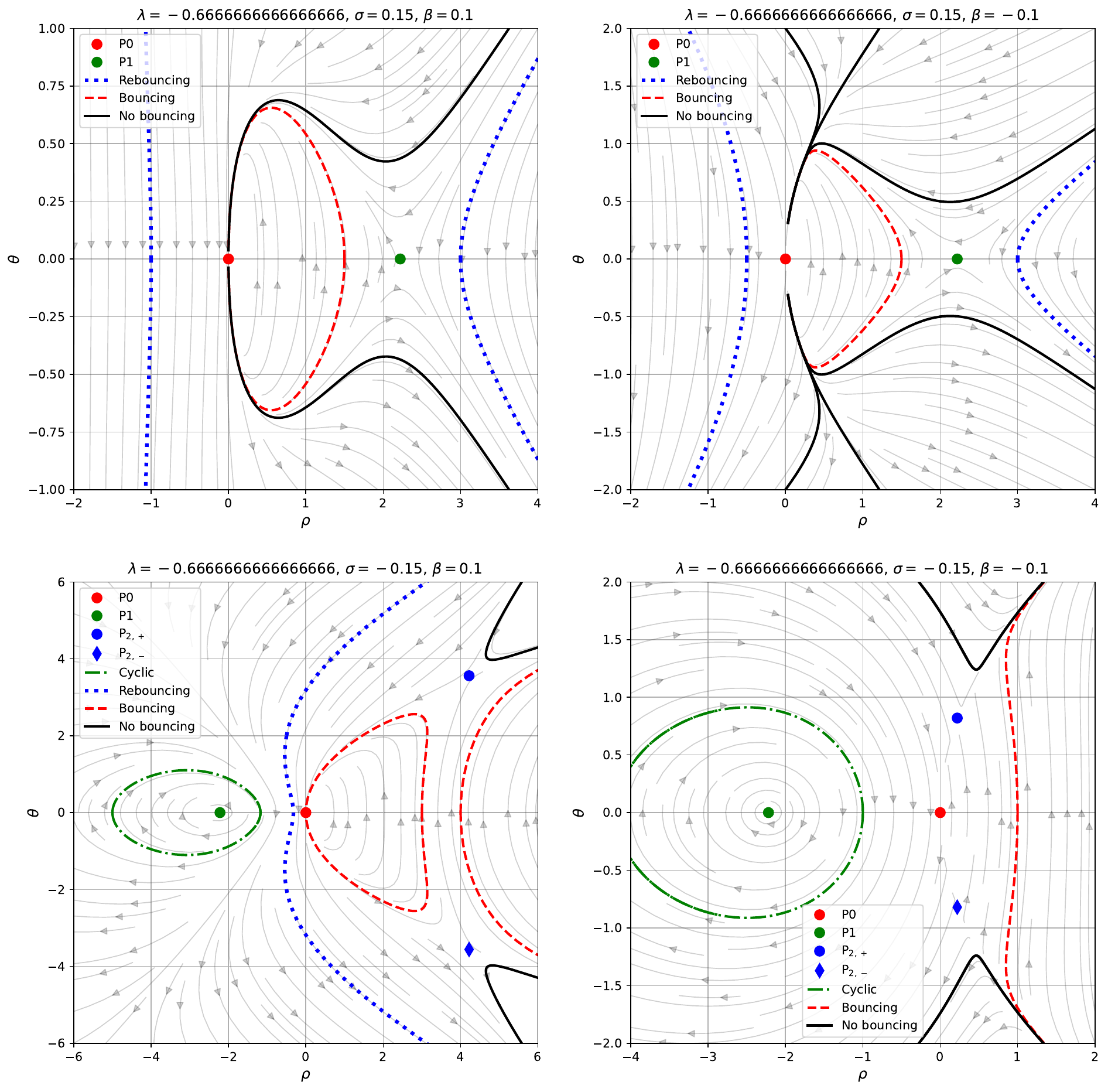} 
\caption{Phase diagrams with $\alpha=0$ and $\lambda=-\frac{2}{3}$. All figures correspond to the combinations $\beta=\pm0.1$ and $\sigma=\pm0.15$. The red dashed curves represent bouncing solutions, the blue dotted curves indicate rebouncing solutions, the green dot-dashed curves correspond to cyclic solutions, and the black solid curves correspond to solutions that do not exhibit a bounce.} 
\label{alpga0lambda2tn} 
\end{figure}

\begin{figure}[h] 
\centering 
\includegraphics[width=1\textwidth]{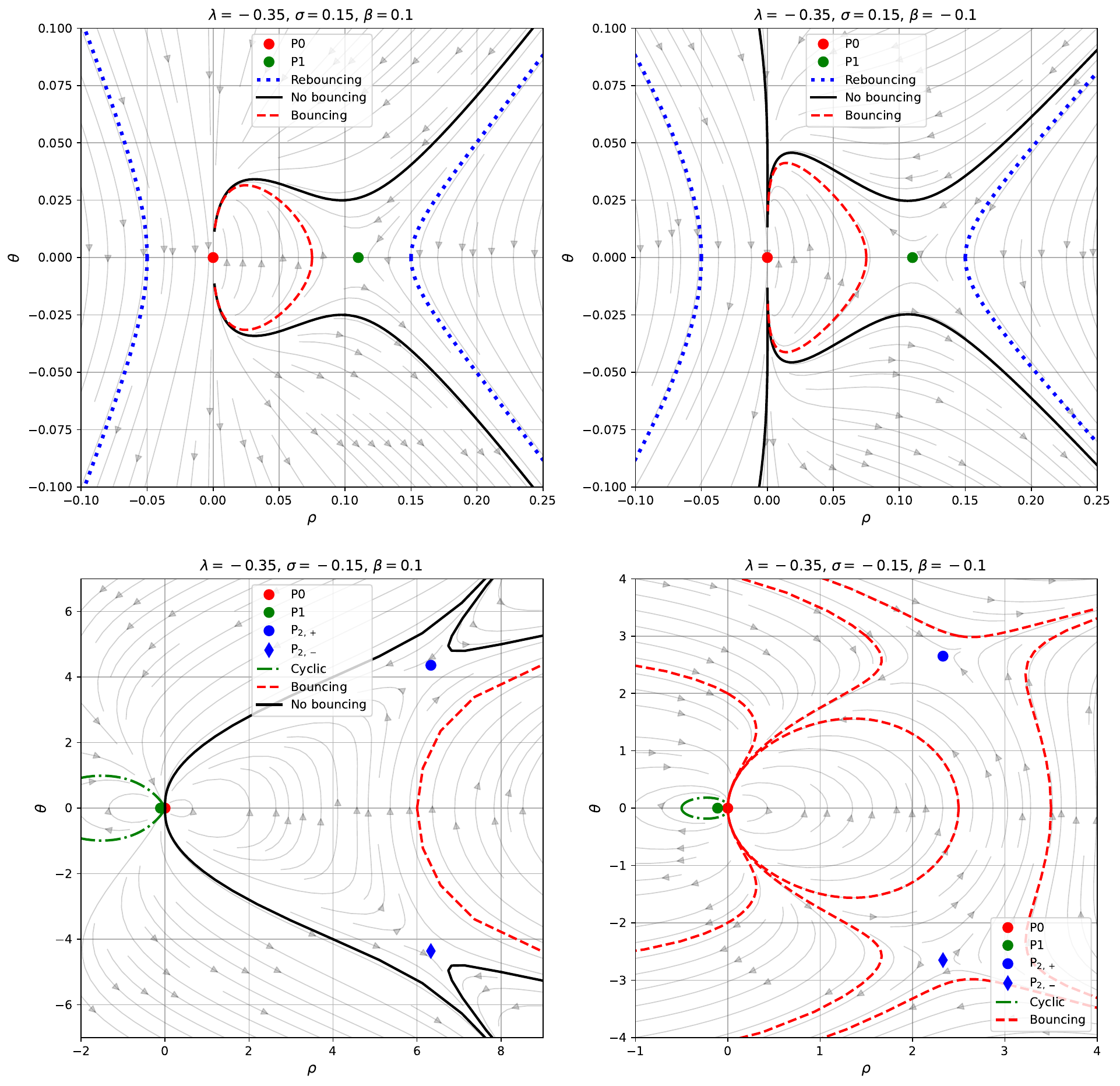} 
\caption{Phase diagrams with $\alpha=0$ and $\lambda=-0.35$. All figures correspond to the combinations $\beta=\pm0.1$ and $\sigma=\pm0.15$. The red dashed curves represent bouncing solutions, the blue dotted curves indicate rebouncing solutions, the green dot-dashed curves correspond to cyclic solutions, and the black solid curves correspond to solutions that do not exhibit a bounce.} 
\label{alpga0lambda0.35n} 
\end{figure}

\begin{figure}[h] 
\centering 
\includegraphics[width=1\textwidth]{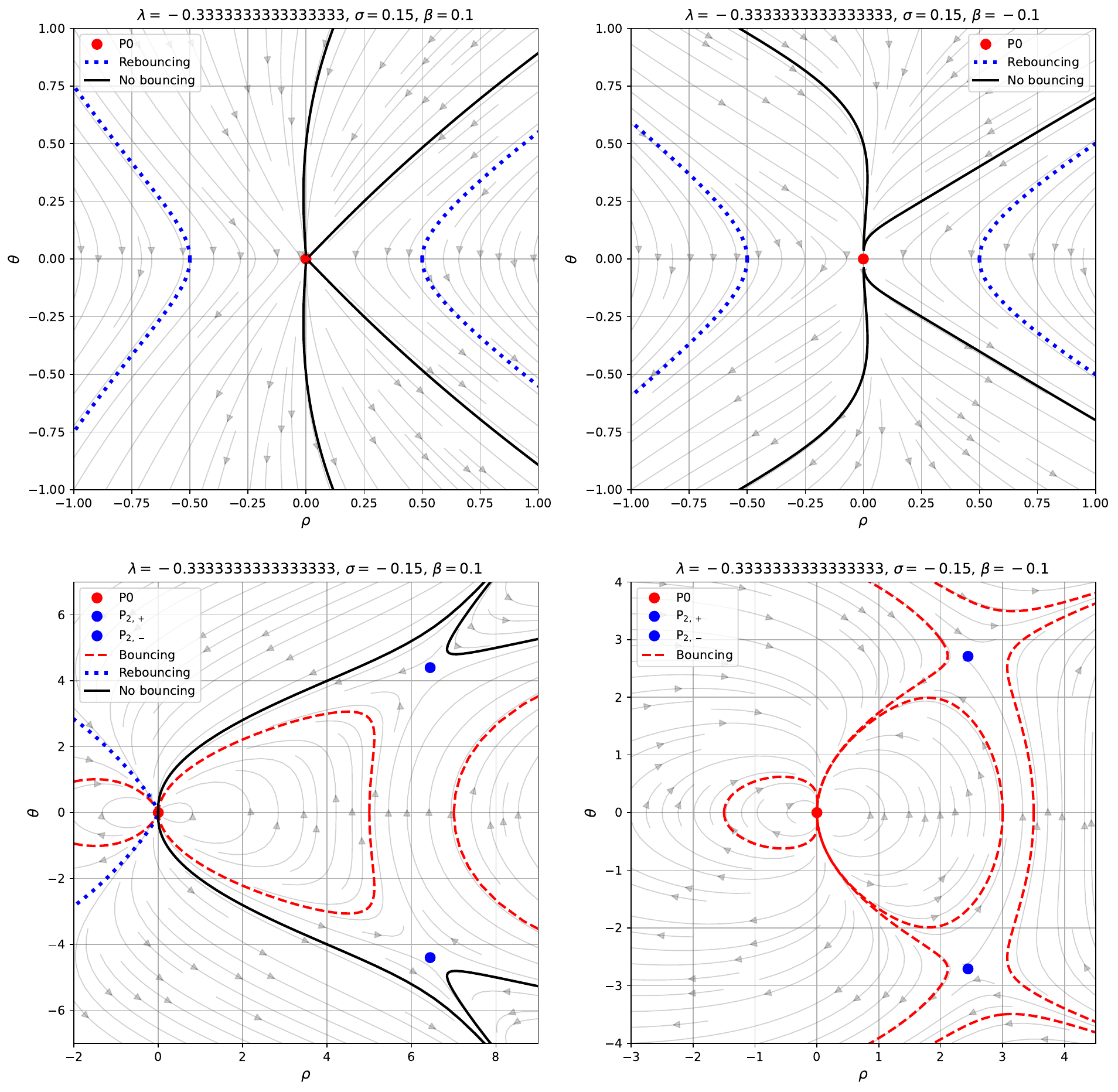} 
\caption{Phase diagrams with $\alpha=0$ and $\lambda=-\frac{1}{3}$. All figures correspond to the combinations $\beta=\pm0.1$ and $\sigma=\pm0.15$. The red dashed curves represent bouncing solutions, the blue dotted curves indicate rebouncing solutions, the green dot-dashed curves correspond to cyclic solutions, and the black solid curves correspond to solutions that do not exhibit a bounce.} 
\label{alpga0lambda1tn} 
\end{figure}

\begin{figure}[h] 
\centering 
\includegraphics[width=1\textwidth]{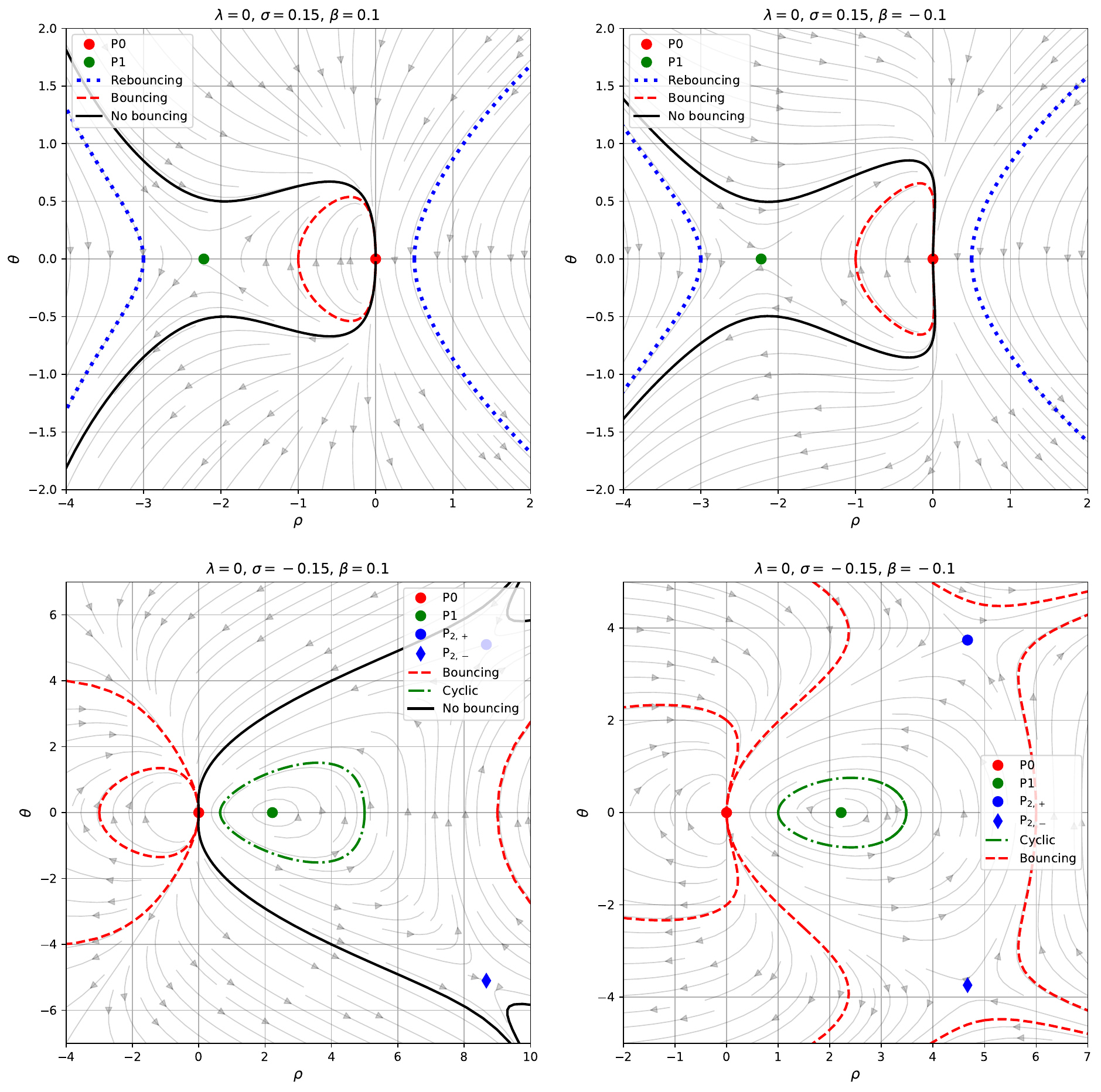} 
\caption{Phase diagrams with $\alpha=0$ and $\lambda=0$. All figures correspond to the combinations $\beta=\pm0.1$ and $\sigma=\pm0.15$. The red dashed curves represent bouncing solutions, the blue dotted curves indicate rebouncing solutions, the green dot-dashed curves correspond to cyclic solutions, and the black solid curves correspond to solutions that do not exhibit a bounce.} 
\label{alpga0lambda0} 
\end{figure}

To understand the sign transition of the $\alpha$ parameter and the subsequent change in stability, it is useful to discuss the model in which the linear correction is absent, i.e., $\alpha=0$. In this case, the equilibria are given by

\begin{equation}
    \label{fixedpointsalpha0}
    P_{0} = (0,0), \qquad P_1 = \left(\frac{-1-3\lambda}{3\sigma}, 0\right),
\end{equation}
and
\begin{equation}
    P_{2,\pm} = \left(-\frac{1+\lambda+3\beta}{\sigma},\ \pm\sqrt{-\frac{3}{\sigma}(1+\lambda+3\beta)}\right).
\end{equation}
Note that $P_{2,\pm}$ only exists when the energy density is positive, i.e., when $\frac{1+\lambda+3\beta}{\sigma}<0$. The qualitative behavior of $P_1$ is exactly the same as discussed in previous sections. $P_{2,\pm}$ is a pair of equilibria: the collapse fixed point ($-$) is always unstable, whereas the expansion fixed point ($+$) can be stable when it behaves as a node. The transition between nodes and saddles occurs when $\sigma$ changes sign. This behavior can be observed in Fig.~\ref{alpga0lambda1n}, \ref{alpga0lambda2tn}, \ref{alpga0lambda0.35n}, \ref{alpga0lambda1tn}, and \ref{alpga0lambda0}.

The main differences are the symmetric behavior of $P_{2,\pm}$, the total absence of the $P_{2,3}$ fixed point, which results in different qualitative behavior for the bouncing and cyclic solutions, and the possibility of a constant rate of collapse/expansion. All the foci and sources now behave as centers, preventing asymptotic behavior for solutions in their vicinity.

In Fig.~\ref{alpga0lambda1n}, the vertical symmetry of the dynamical system becomes evident. This symmetry prohibits the bouncing solutions, delimited by the saddle points, from escaping into eternal collapse or solutions with accelerated expansion factor as seen in the last section. It is also noticeable that the focus/source behavior seen in Fig~\ref{gerallambda1n} now gives way to a center; the rest of the behavior is analogous. Much of the discussion follows similarly: the symmetry of the phase diagrams makes the different regions more clearly delimited, and solutions remain within the regions where they originate.

\begin{figure}[h] 
\centering 
\includegraphics[width=1\textwidth]{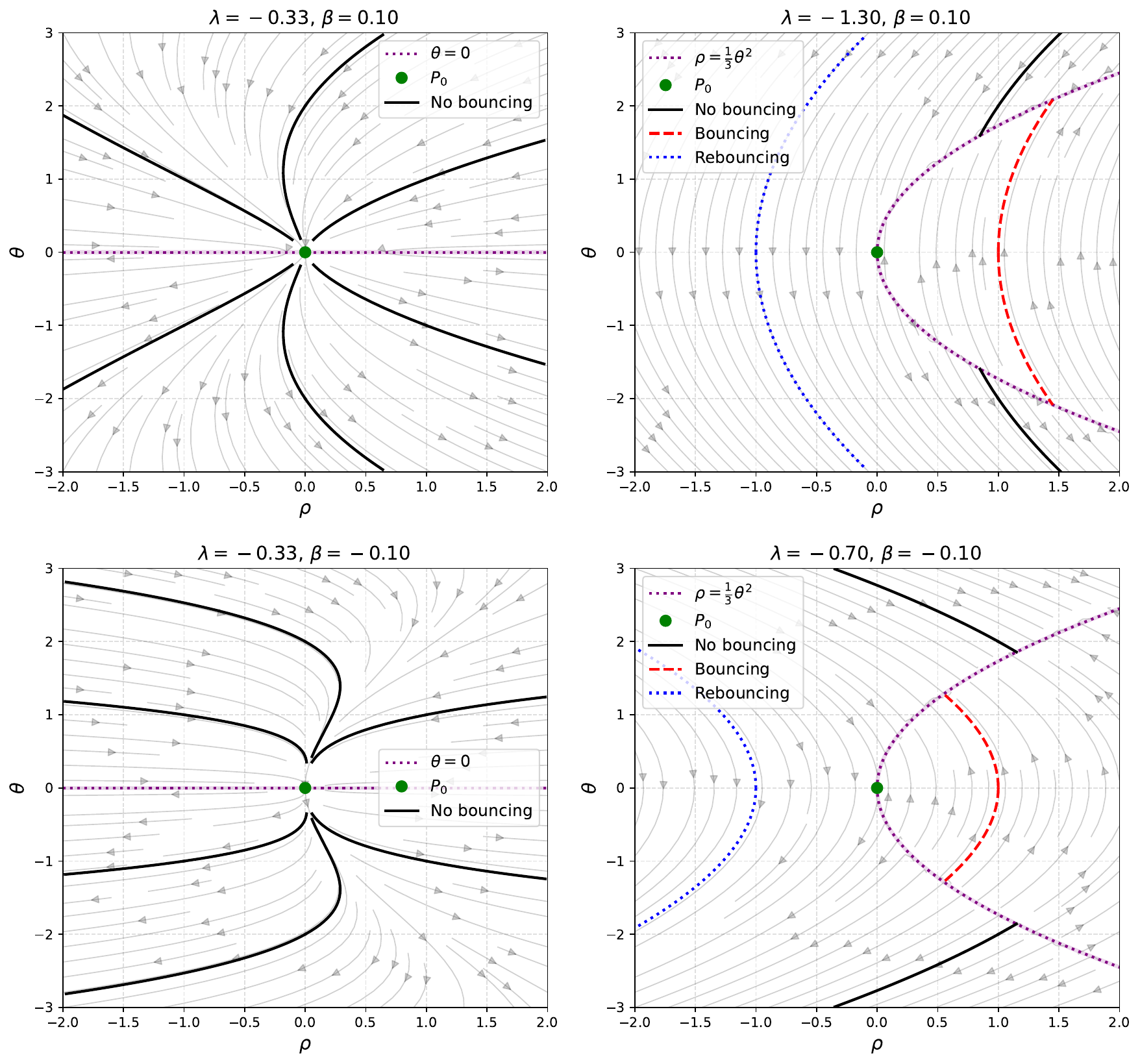} 
\caption{Phase diagrams with $\alpha=0$ and $\sigma=0$, presenting the particular cases where $\lambda=-\frac{1}{3}$ (left) and $1+\lambda+3\beta=0$ (right). All figures correspond to the combinations $\beta=\pm0.1$. The red dashed curves represent bouncing solutions, the blue dotted curves indicate rebouncing solutions, and the black solid curves correspond to solutions that do not exhibit a bounce.} 
\label{alpga0sigma0special} 
\end{figure}

In particular, when $\sigma=0$ in this situation, there are no fixed points, but there are tendency curves. If $\lambda=-\frac{1}{3}$, the entire line $\dot{\theta}=0$ is a tendency curve, and for $1+\lambda+3\beta=0$, there is a tendency parabola given by $\theta^2= 3\rho$; this can be observed in Fig.~\ref{alpga0sigma0special}.

The behavior shown in Fig.~\ref{alpga0sigma0special} is somewhat curious. Solutions that start on the line $\theta=0$ never leave that curve, but small negative $\theta$ can lead to eternal collapse, just as solutions with positive $\theta$ always approach the origin. This behavior results from the merging of the saddles and nodes given by $P_1$ and $P_{2,i}$, which degenerate at the origin for certain parameter choices. More interestingly, when $1+\lambda+3\beta=0$, a universe with the usual relation $\rho = \frac{\theta^2}{3}$ - usually obtained from a null spatial curvature in GR - is unstable for collapse initial conditions, collapsing forever, or bouncing back into a stable state of the same relation, but with a positive expansion factor. 

To analytically discuss the qualitative behavior of the origin, it is necessary to study higher-order corrections of the dynamical system and use different techniques than those used in this section. The next subsection will focus on using the center manifold theorem (\cite{kuznetsov2004}) to discuss the stability of the origin equilibrium.

\subsection{Stability analysis of the non-hyperbolic origin}

\begin{figure}[h] 
\centering 
\includegraphics[width=1\textwidth]{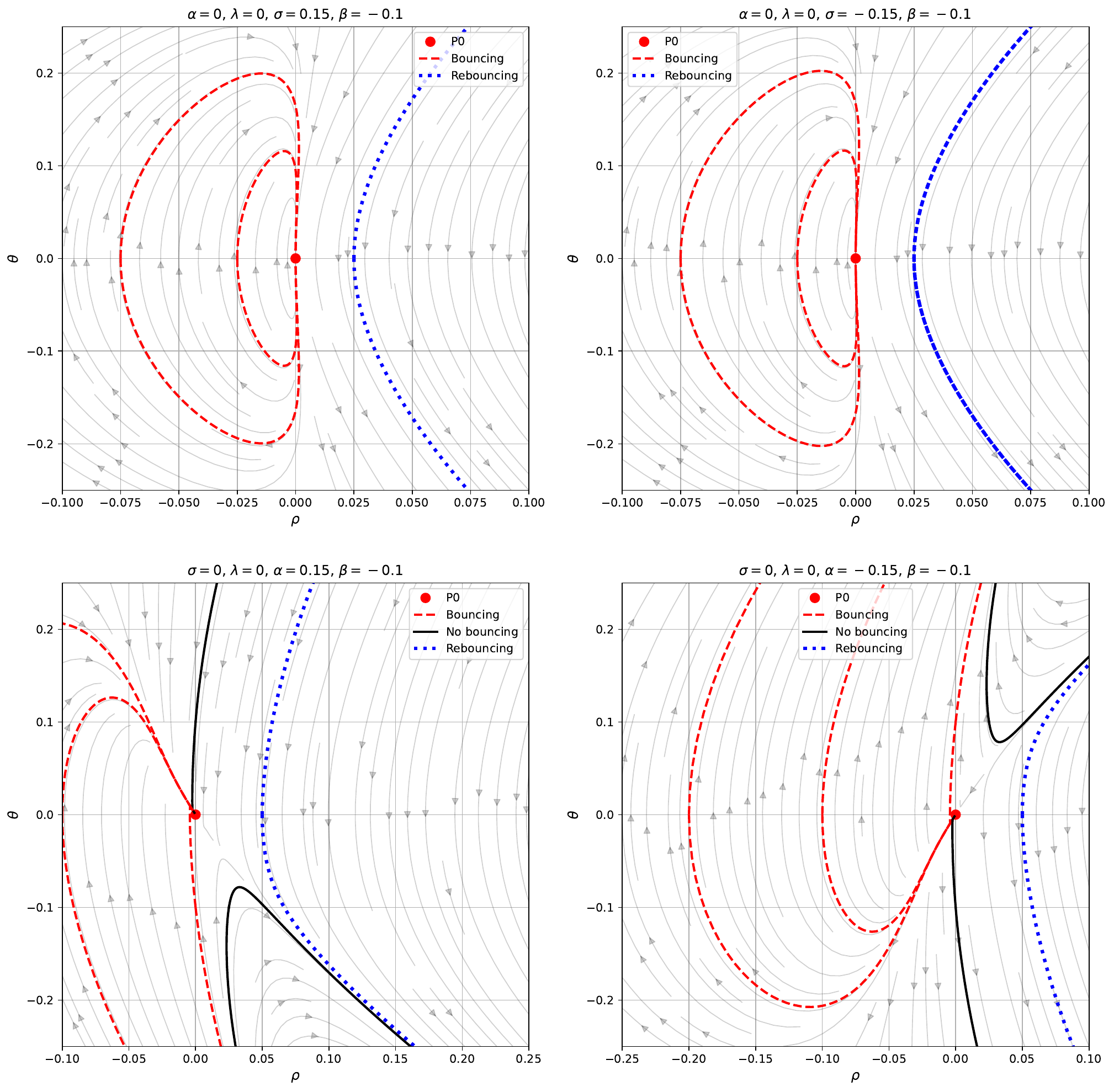} 
\caption{Phase diagrams with $\alpha=0$ (top row) for $\sigma=\pm0.15$. Also for $\sigma=0$ (bottom row) for $\alpha=-0.15$. All graphs are made for $\lambda=0$ and $\beta=-0.1$, zoomed to observe the behavior near the origin. The red dashed curves represent bouncing solutions, the blue dotted curves indicate rebouncing solutions, and the black solid curves correspond to solutions that do not exhibit a bounce.} 
\label{alpga0sigma0specialzoom} 
\end{figure}

For this discussion, we will use the center manifold theorem. The theorem requires that we construct a family of curves that obey the expression $\theta = h(\rho)$, where $h(0) = 0$ and $h'(0) = 0$, with $h'(\rho) = \frac{dh}{d\rho}$. This allows us to reduce the system to a first-order equation near the origin and discuss the qualitative behavior through this reduced equation. We consider a second-order expansion of the form
\begin{equation}
\label{centermanifold}
    h(\rho) = h_2 \rho^2 + \mathcal{O}(\rho^3),
\end{equation}
where $h_2$ is a parameter to be determined using equations (\ref{generalds}). Notice that the zeroth- and first-order terms vanish since the function must pass through the origin and be tangent to it. The relation we obtain is of the form
\begin{equation}
\label{coefficients}
    0 = -\frac{1}{2}(1+3\lambda)\rho-\frac{3}{2}(\sigma+h_2\alpha)\rho^2.
\end{equation}
The expression (\ref{coefficients}) was taken up to second order in $\rho$. Taking $\alpha \ne 0$, we find that $h_2 = -\frac{\sigma}{\alpha}$. The linear term is due to the form of the dynamical system and does not affect the approximation used in the calculation of $h(\rho)$. It is important to notice that for $\lambda = -\frac{1}{3}$, the dominant term is the quadratic one. Using (\ref{centermanifold}) inside the equation for $\dot{\rho}$, we obtain
\begin{equation}
    \label{reducedcontinuity}
    \dot{\rho} = \frac{\sigma}{\alpha}(1+\lambda)\rho^3, \qquad h(\rho) = -\frac{\sigma}{\alpha}\rho^2.
\end{equation}
Equation (\ref{reducedcontinuity}) gives us information about the direction in which the trajectories move in each of the quadrants near the origin. For example, if $\theta < 0$ and $\frac{\sigma}{\alpha} > 0$, all solutions will move away from the origin, but for $\theta > 0$ the same choice of parameters will guarantee that the solutions move towards the origin. If $\sigma = 0$ or $\alpha = 0$, there is too much degeneracy to be studied through the center manifold theorem; the second-order expansion will not suffice, and higher-order expansions yield either trivial solutions for some orders or constrain the parameters of the model instead of the expansion parameters, leaving them undefined. Increasing the order of the expansion does not change this behavior. Nonetheless, the behavior for the general case gives us some insight into the behavior of these cases, allowing us to discuss them with reference to Figure~\ref{alpga0sigma0specialzoom}.

As seen in the top row of Fig.~\ref{alpga0sigma0specialzoom}, the degeneracy of $P_1$ with the origin when $\sigma = 0$ causes the loop that results in the bouncing solutions near the origin, originating from the static universe and bouncing back into this same state. The rebouncing solutions can be tangent to the origin, after which they collapse forever. For $\alpha = 0$, we can see that changing the sign of $\sigma$ can change the stability of the origin. Now, since $P_1$ is not degenerate with the origin, not all solutions bounce into a static universe, but may instead collapse forever when $\alpha > 0$, or evolve towards a universe with positive expansion factor when $\alpha < 0$ due to the stability of $P_1$.
 
\section{Conclusions}

We studied the addition of second-order terms in the EoS of a cosmological model whose geometry obeys GR, and presented a qualitative analysis of its behavior under different choices of the model's free parameters: $\lambda$, $\sigma$, $\alpha$, and $\beta$. By varying these parameters, we found the presence of bifurcations in the dynamical system, with solutions representing different cosmic configurations. It is possible to observe bouncing, rebouncing, cyclic models, as well as solutions that collapse or expand forever.

The stability of equilibrium points provides several relevant insights regarding the final state of possible cosmic scenarios. We note that expanding universes present stable equilibria, while collapsing ones are unstable. Bouncing solutions can have several final configurations: it is possible to obtain bouncing solutions that end at a constant rate in expanding universes, others whose expansion is accelerated, or even bounces that result in a Minkowski universe (static with zero energy density). Moreover, we show that there are solutions in which an increase in the size of the universe does not necessarily imply that the energy density decreases with expansion.

We observe that $\sigma$ and $\alpha$ mainly controls the stability and behavior of $P_0$ and $P_1$ equilibria. The parameter $\alpha$ is also responsible for the asymmetry of the solutions and affects the stability of the equilibria $P_{2,i}$ that appear due to the expansions in $\theta$. The most relevant effect of $\beta$ is that it allows more freedom in fixing $\lambda$ while still obtaining different qualitative behaviors, and it increases the number of bouncing solutions in situations where $\sigma < 0$. For future work, it is expected that constraints on the parameters through observational data should be established, allowing us to study how these viscosity corrections could influence structure formation or the inflationary epoch, especially in cases where the final state of the universe presents a constant expansion rate, as this appears to be the current state of our universe.

\section*{Acknowledgments}
The authors thank the support of CAPES (AGC grant N. 88887.666979/2022-00), and FAPERJ (MN is Emeritus Visiting Researcher fellow).

\bibliography{ref}

\end{document}